# Realization of multiple topological states and topological phase transitions in (4,0) carbon nanotube derivatives

Yan Gao[1]*, Yu Du[1#], Yun-Yun Bai[1#], Weikang Wu[2], Qiang Wang[1], Yong Liu[1]*, Kai Liu[3]*, and Zhong-Yi Lu[3]*

[1]*State Key Laboratory of Metastable Materials Science and Technology and Key Laboratory for Microstructural Material Physics of Hebei Province, School of Science, Yanshan University, Qinhuangdao 066004, China*

[2]*Key Laboratory for Liquid-Solid Structural Evolution and Processing of Materials (Ministry of Education), Shandong University, Jinan, Shandong 250061, China*

[3]*Department of Physics and Beijing Key Laboratory of Opto-electronic Functional Materials & Micro-nano Devices, Renmin University of China, Beijing 100872, China*

## Abstract

Exploring various topological states (TS) and topological phase transitions (TPT) has attracted great attention in condensed matter physics. However, so far, there is rarely a typical material system that can be used as a platform to study the TS and TPT as the system transforms from one-dimensional (1D) nanoribbons to two-dimensional (2D) sheet then to three-dimensional (3D) bulk. Here, we first propose that some typical TS in 1D, 2D, and 3D systems can be realized in a tight-binding (TB) model. Following the TB model and further based on first-principles electronic structure calculations, we demonstrate that the structurally stable (4,0) carbon nanotube derivatives are an ideal platform to explore the semiconductor/nodal-point semimetal states in 1D nanoribbons [1D-(4,0)-$C_{16}H_4$ and 1D-(4,0)-$C_{32}H_4$], nodal-ring semimetal state in 2D sheet [2D-(4,0)-$C_{16}$], and nodal-cage semimetal state in 3D bulk [3D-(4,0)-$C_{16}$]. Furthermore, we calculate the characteristic band structures and the edge/surface states of 2D-(4,0)-$C_{16}$ and 3D-(4,0)-$C_{16}$ to confirm their nontrivial topological properties. Our work not only provides new excellent 2D and 3D members for the topological carbon material family, but also serves as an ideal template for the study of TS and TPT with the change of system dimension.

[#] Y. Du and Y.-Y. Yun contributed equally to this work.

*Corresponding authors: yangao9419@ysu.edu.cn; yongliu@ysu.edu.cn; kliu@ruc.edu.cn; zlu@ruc.edu.cn

# Introduction

Topological semimetals (TSMs) have attracted widespread interest in condensed matter physics and materials science due to a variety of emergent particles and interesting quantum transport properties as well as their potential future applications[1-7]. TSMs can be classified based on the key attributes of the band crossing points near the Fermi level, such as their degeneracy, dispersion, and distribution in the momentum space[5]. According to these attributes, various types of TSM families have been theoretically proposed and experimentally confirmed, including Dirac semimetals[8-10], Weyl semimetals[11-15], multifold fermion semimetals[16-18], type-I and type-II semimetals[19,20], nodal line semimetals[21-23], and so on[24-29]. Most of these topological phases focus on materials containing heavier elements[4,5], which are expected to give rise to nontrivial band topology through strong spin-orbit coupling (SOC) effect. Instead, the materials containing light elements seem to offer an easier alternative to capturing TSMs[30-34], because the SOC effect in such systems is inherently weak enough to be neglected, and TSMs are very likely to be obtained once the energy bands cross each other near the Fermi level.

Actually, many TSM states have been theoretically proposed in 2D and 3D carbon allotropes[33-40]. On the one hand, carbon has diverse $sp$, $sp^2$, and $sp^3$ hybridization states that can form a variety of carbon allotropes; On the other hand, the metallic behavior of carbon allotropes usually originates from the graphene-like $p_z$ orbitals of $sp^2$ carbon atoms[33], which provides directional guidance for designing TSM states in carbon materials. However, until now, few materials have served as a typical example of simultaneous realization of multiple topological states and topological phase transitions (TPT) with the change of system dimensions. It is worth noting that 3D carbon networks connected by 1D carbon nanotubes (CNT) have attracted much attention due to their potential applications in nanoelectronics and nanomechanics[41,42]. In addition, the 3D carbon network composed of (3,3) single-wall CNT can realize topological nodal-line and nodal-surface semimetal states[43]. Particularly, a series of 3D carbon networks consisting of (4,0)-CNT and armchair graphene nanoribbons have recently been proposed as an ideal platform for the realization of nodal flexible-surface semimetals[44]. Therefore, a natural and interesting question is whether a variety of typical topological states and TPT can be realized in the 3D (4,0)-CNT derivatives by using only the 1D (4,0)-CNT as the building block.

In this work, we first propose the trivial insulator, topological nodal-point, nodal-ring, and nodal-cage semimetal states in the TB model of eight sites consisting of the upper and lower square layers. Following the guidance of the TB model, using 1D (4,0)-CNT as the building block, we find that a series of (4,0)-CNT derivatives [such as 1D-(4,0)-$C_{16}H_4$, 1D-(4,0)-$C_{32}H_4$, 2D-(4,0)-$C_{16}$, and 3D-(4,0)-$C_{16}$] can serve as a typical example to realize these trivial/nontrivial topological states and TPT. The phonon spectra, *ab initio* molecular dynamics simulations, and the elastic constants calculations confirm the dynamical, thermal dynamical (above 600 K), and mechanical stabilities of these (4,0)-CNT derivatives, respectively. We show their characteristic band structures (including nodal point, nodal ring, and nodal cage) by first-principles electronic structure calculations. The existences of the edge states of 2D-(4,0)-$C_{16}$ and the surface states of 3D-(4,0)-$C_{16}$ further confirm their topological nontrivial properties.

## Computational Methods

The electronic structures of the (4,0)-CNT derivatives were studied with the projector augmented wave method[45] as implemented in the VASP code[46] based on density functional theory (DFT). The generalized gradient approximation (GGA) described by the Perdew–Burke–Ernzerhof (PBE) type[47] was adopted for the exchange–correlation functional. The $10 \times 1 \times 1$, $10 \times 10 \times 1$, and $10 \times 10 \times 12$ *k*-point meshes[48] were taken for the Brillouin zone (BZ) integration of the 1D nanoribbons [1D-(4,0)-$C_{16}H_4$ and 1D-(4,0)-$C_{32}H_4$], 2D-(4,0)-$C_{16}$, and 3D-(4,0)-$C_{16}$, respectively. The vacuum layer of 20 Å was applied to avoid the residual interactions between the neighboring images for the 1D nanoribbons and 2D sheet calculations. The convergence criteria for energy and force were set to $10^{-6}$ eV and 0.001 eV/Å, respectively. The phonon spectrum calculations were performed by using the first-principles quantum transport package Nanodcal[49]. The thermal stability was studied with the *ab initio* molecular dynamics (AIMD) simulations in a canonical ensemble with a Nose-Hoover thermostat[50]. The topological properties of (4,0)-CNT derivatives were investigated by using the Wannier90[51] and WannierTools[52] packages.

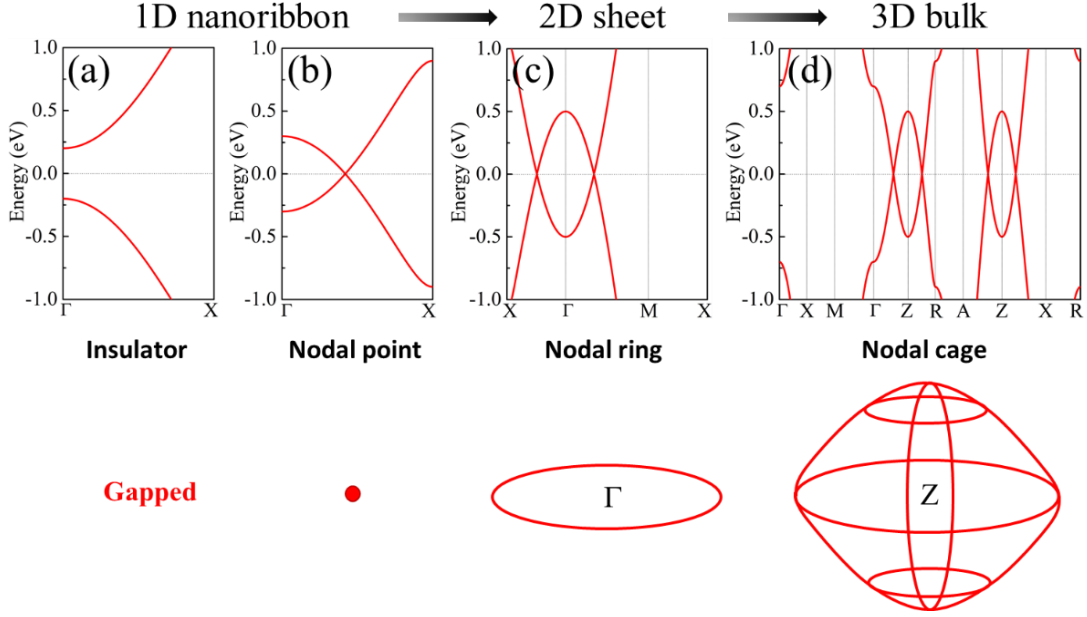

FIG. 1. (Color online) Typical topological phase transitions with the change of system dimensions. Evolution from (a) direct band-gap semiconductor to (b) nodal-point semimetal in 1D nanoribbon to (c) nodal-ring semimetal in 2D sheet to (d) nodal-cage semimetal in 3D bulk. The lower panels of (a-d) correspond to the insulator state, topological nodal point, nodal ring, and nodal cage state, respectively.

## Results and Discussions

To investigate these typical topological states and topological phase transitions (TPT) [see Fig. 1], we construct a tetragonal tight-binding (TB) model of eight sites consisting of the upper and lower square layers [see Fig. 2]. Since the spin-orbit coupling (SOC) effect of carbon is so weak as to be negligible, the TB Hamiltonian with only one orbital at each site can be expressed as:

$$H = \sum_{i,j} \sum_{\mu} t_{ij} e^{-i k r_{ij}^{\mu}}, \qquad (1)$$

where $i$, $j$ label the eight sites, $\mathbf{k}$ is the reciprocal lattice vector, $\mathbf{r}_{ij}^{\mu}$ is the distance vector directed from site $j$ to site $i$, $t_{ij}$ is the hopping energy (in units of eV) between sites $i$ and $j$, and $\mu$ runs over all equivalent hopping sites. The hopping parameters ($t_{ij}$) within the primitive cell are $t_1$ and $t_2$, and the hopping parameters between the primitive cell are $t_{2a}$, $t_{2b}$, and $t_3$ [see Fig. 2]. It is clear that we can adjust these parameters to simulate the topological trivial/nontrivial state that occurs in different dimensional cases. For example, to describe the evolution of the system from 1D nanoribbons to 3D bulk, we only need to add $t_{2b}$ (2D case) and $t_{2b}$, $t_3$ (3D case) on the basis of hopping parameters $t_1$, $t_2$, and $t_{2a}$ (1D case). Therefore, we can set $t_1 = 2.8$, $t_2 = 0.9$, and $t_{2a} = 0.6$ to obtain the insulator state in 1D nanoribbons [see Fig. 1(a)]. With the increase of $t_{2b}$, the conduction band minimum (CBM) moves down and the valence band maximum (VBM) moves up, thereby the two bands are reversed at the $\Gamma$ point, resulting in a nodal-point semimetal [see Fig. 1(b)]. Once $t_{2b} = t_{2a} = 0.6$, the system evolves into a 2D sheet and a nodal-ring semimetal is obtained [see Fig. 1(c)]. If $t_3$ is further increased ($t_3 = 0.5$), the system expands along the $c$-axis into a 3D bulk, and we can get a nodal-cage semimetal [see Fig. 1(d)].

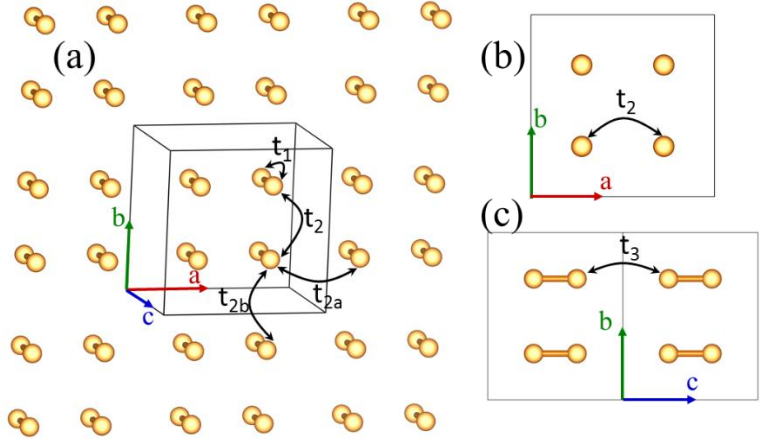

FIG. 2. (Color online) The tight-binding (TB) model for topological phase transitions. (a) Perspective, (b) top view, and (c) side view of the TB model of eight sites consisting of the upper and lower square layers, where the $c$-axis and $a$/$b$ axes represent the directions of the square tube and the inter-tube directions, respectively. The black box represents the primitive

cell of the TB model. The intra-tube hoppings considered are $t_1$, $t_2$, and $t_3$, and the inter-tube hoppings are $t_{2a}$ and $t_{2b}$, respectively.

Next, we consider the realizations of these multiple topological states (nodal point/ring/cage) and TPT in real materials. Following the guidance of the TB model, when there is only one orbital at each site and the SOC effect can be neglected, we naturally think of carbon material because of its light weight and diversity of allotrope structures. In addition, its metallic properties are mainly determined by the $p_z$ or similar $p_z$ orbital of the *sp*² carbon atoms. Therefore, we choose carbon allotropes to achieve these typical topological states. We find that (4,0)-CNT can be used as the basic unit of our model, because (4,0)-CNT may form a 3D carbon network structure [namely 3D-(4,0)-$C_{16}$] by squeezing each other [see Fig. 3(h)]. If the 8 carbon atoms (C1) of the *sp*³ type bonding between nanotubes are removed, the TB model proposed by us will be naturally satisfied. Based on 3D-(4,0)-$C_{16}$, we can cut a film with a thickness of four atomic layers along the direction perpendicular to 1D (4,0)-CNT to obtain the 2D sheet (namely 2D-(4,0)-$C_{16}$) [see Figs. 3(e-f)], and further cut the 2D sheet along the direction perpendicular to the bonding between the nanotubes with different layers of thickness [n, shown in the inset of Fig. 3(a)], and then obtain a 1D nanoribbons with widths (n) of 1 and 2 (4,0)-CNT after passivation with hydrogen [see Figs. 3(a-d)].

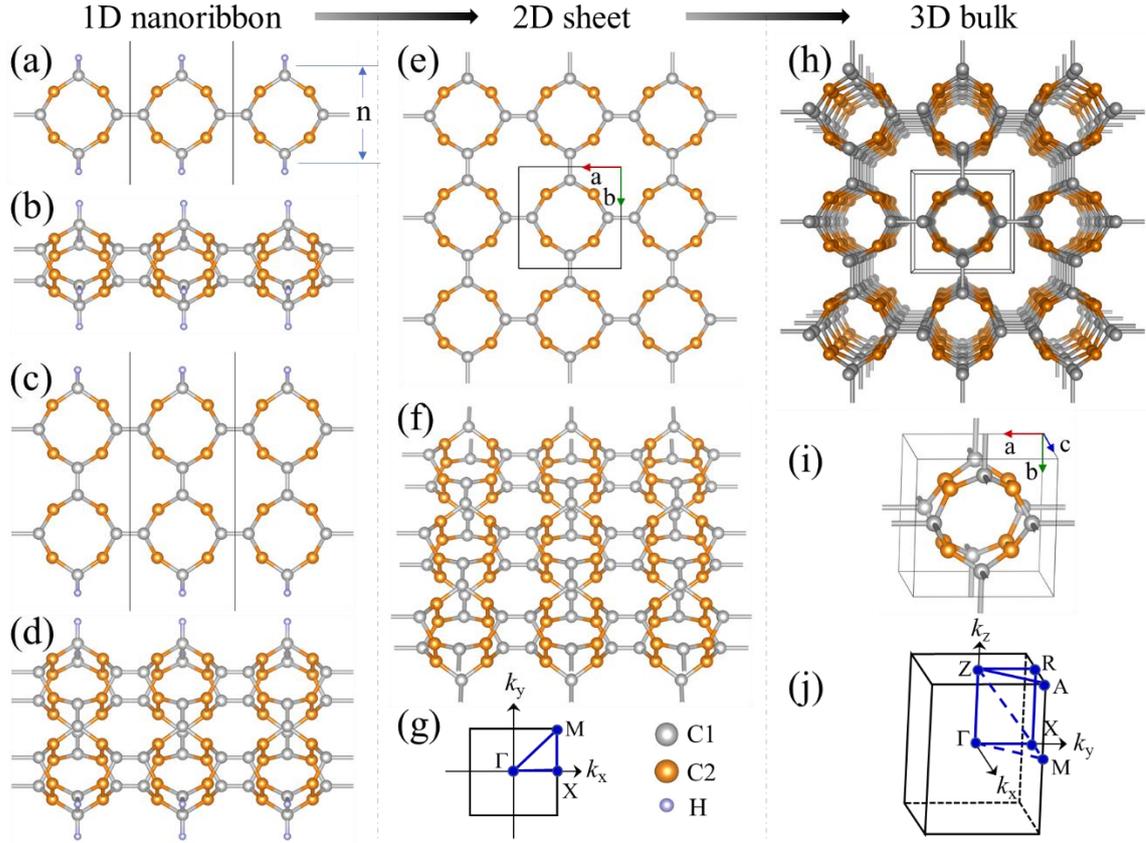

FIG. 3. (Color online) Evolution of (4,0)-CNT derivatives from 1D nanoribbon to 3D bulk structures. The 1D nanoribbons with widths (n) of 1 and 2 (4,0)-CNT film after hydrogen passivation, namely (a-b) 1D-(4,0)-$C_{16}H_4$ and (c-d) 1D-(4,0)-$C_{32}H_4$. (e) Top view and (f) perspective view of the 2D (4,0)-CNT sheet [2D-(4,0)-$C_{16}$]. (g) Brillouin zone (BZ) of the 2D-(4,0)-$C_{16}$. (h) Perspective view and (i) primitive cell of 3D bulk [3D-(4,0)-$C_{16}$]. (j) BZ of the 3D-(4,0)-$C_{16}$.

In order to verify the stability of these (4,0)-CNT derivatives, we first calculated the phonon spectra of 1D-(4,0)-$C_{16}H_4$, 1D-(4,0)-$C_{32}H_4$, 2D-(4,0)-$C_{16}$, and 3D-(4,0)-$C_{16}$, as shown in Figs. S1(a) and S1(b) of Supplemental Material (SM), Fig. 5(a), and Fig. S2 in the SM, respectively. Clearly, these phonon spectra have no imaginary phonon mode in the entire Brillouin zone (BZ), thus these crystal structures are dynamically stable. We also examine the thermal stability of 2D-(4,0)-$C_{16}$ and 3D-(4,0)-$C_{16}$ by the AIMD simulations, showing they can be stable under at least 600 K (see Fig. 5(b) and Fig. S3 in the SM). The calculated total energies of 3D-(4,0)-$C_{16}$ and 2D-(4,0)-$C_{16}$ are -8.61 eV and -8.28 eV per atom, respectively, which are 0.70 eV and 0.37 eV/atom lower than experimentally realized T-carbon[53,54] (−7.91 eV/atom). Besides, we calculated the elastic constants of 3D-(4,0)-$C_{16}$ and 2D-(4,0)-$C_{16}$ to assess their mechanical stability. For 3D-(4,0)-$C_{16}$, the six independent elastic constants $C_{11}$, $C_{12}$, $C_{13}$, $C_{33}$, $C_{44}$, and $C_{66}$ were computed to be 437.5, 175.4, 76.5, 888.1, 220.4, and 144.7 GPa, respectively. Obviously, they fulfill the criteria of mechanical stability for tetragonal (I) phase[55]: $C_{11} > |C_{12}|$, $2C_{13}^2 < C_{33}(C_{11} + C_{12})$, $C_{44} > 0$, and $C_{66} > 0$. For 2D-(4,0)-$C_{16}$ in the square lattice, the calculated values of three independent elastic constants $C_{11} = C_{22}$, $C_{12}$, and $C_{66}$ are 234.6, 104.7, and 88.1 N/m, respectively, which satisfy the stability criteria[55] $C_{11}C_{22} - C_{12}^2 > 0$ and $C_{66} > 0$. Therefore, both 3D-(4,0)-$C_{16}$ and 2D-(4,0)-$C_{16}$ are mechanically stable as well. We further obtain the in-plane Young's modulus [see Fig. 5(c)] and Poisson's ratio [see Fig. 5(d)] of 2D-(4,0)-$C_{16}$ as functions of the angle with respect to the $a$-axis of the primitive cell, from which we can see that both $Y_{(\theta)}$ and $\nu_{(\theta)}$ have obvious deviation from the circular behavior, showing moderate anisotropy.

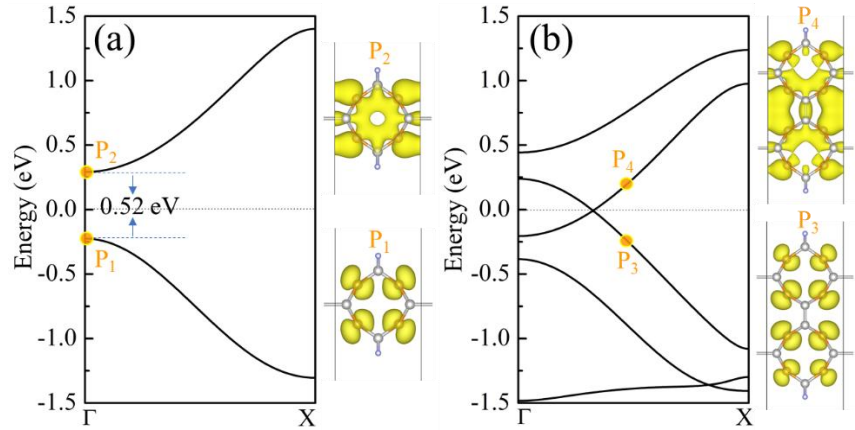

FIG. 4. (Color online) Band structures and charge densities of the 1D-(4,0)-$C_{16}H_4$ and 1D-(4,0)-$C_{32}H_4$.

The most interesting property in the (4,0)-CNT derivatives is the topological phase transitions from a topological trivial insulating state to a nontrivial nodal-point semimetal then to a nodal-ring semimetal and to a nodal-cage semimetal through the adjustment of the system dimensions. Figure 4(a) shows the band structure of a (4,0)-CNT film of width 1 [see Fig. 3(a)] after hydrogen passivation (namely 1D-(4,0)-$C_{16}H_4$), from which we can see that it is a semiconductor with a direct band gap of 0.52 eV. Once the width (n) change to 2 [see Figs. 3(c-d)], the CBM of 1D-(4,0)-$C_{16}H_4$ shifts down and the VBM shifts up due to the increased interaction between carbon atoms in the *b*-axis direction, resulting in band inversion and forming a nodal-point semimetal. In addition, we can see from their charge density near the Fermi level that it is mainly from the C2 atom's graphene-like $p_z$ orbital contribution, which is consistent with our TB model.

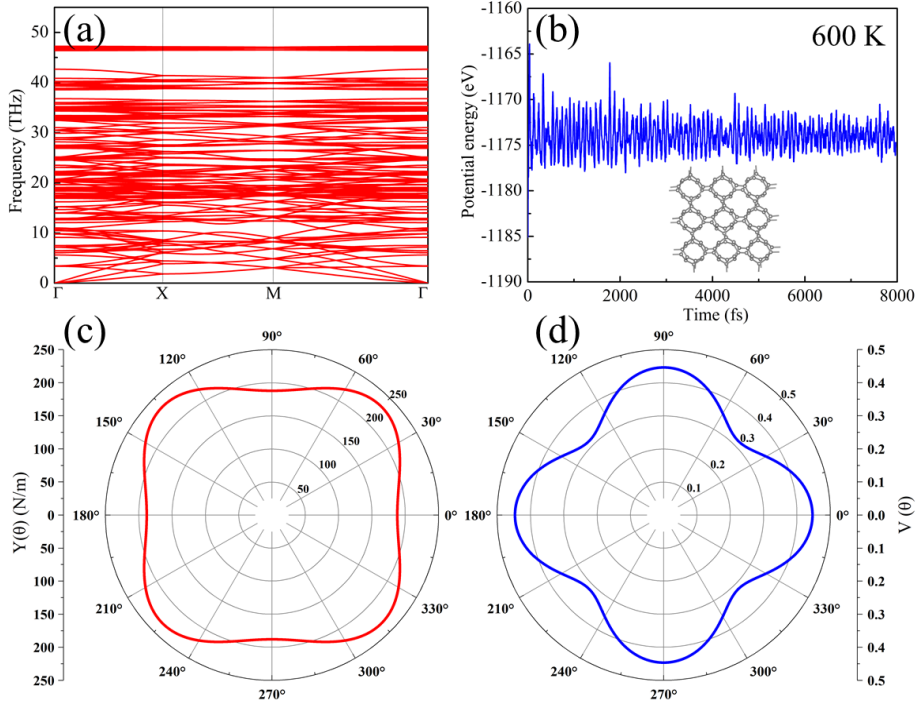

FIG. 5. (Color online) Dynamical, thermal, and mechanical stabilities of 2D-(4,0)-$C_{16}$. (a) Phonon dispersion of 2D-(4,0)-$C_{16}$ in the whole BZ. (b) The total potential energy fluctuation of 2D-(4,0)-$C_{16}$ during the AIMD simulation at 600 K. The inset is the equilibrium structure of 2D-(4,0)-C16 after 8000 fs of AIMD simulation at 600 K. Polar diagrams for Y(θ) (c) and ν (θ) (d) of 2D-(4,0)-$C_{16}$.

As the width n further increases, a 2D structure consisting of a (4,0)-CNT film with four atomic layer thick is constructed, i.e. 2D-(4,0)-$C_{16}$ [see Figs. 3(e-f)]. We calculated its band structure, as shown in Fig. 6(a). One can see that 2D-(4,0)-$C_{16}$ has only two interpenetrating energy bands near the Fermi level to form a perfect nodal ring around the Γ point [see Figs. 6(b-c)]. It is found by calculation that the two intersecting bands have opposite eigenvalues of the horizontal mirror [see Fig. 6(a)], so the nodal ring is protected by the mirror symmetry[56]. To confirm this fact, we also calculated the bands of 2D-(4,0)-$C_{16}$ under horizontal biaxial and uniaxial strains (see Figs. 6(e-h) and Fig. S4 in the SM), and their nodal rings still exist. Moreover, we can clearly see from the charge densities of the four points (a, b, c, and d) in Fig. 6(a) that the charge densities at points a and c are similar, while the charge densities at points b and d are similar [see Fig. 6(d)], suggesting that there is a band inversion near the intersection. Notably, these charge densities near the Fermi level are derived from the graphene-like $p_z$ orbital of the C2 atom, which is in good agreement with our TB model. In order to further confirm its nontrivial topological properties, we calculated the edge states of 2D-(4,0)-$C_{16}$ along the [100] direction [see Fig. 6(i)], from which we can see that the nodal ring has obvious characteristics of the drumhead-like state, and the edge state is nestled outside the projection nodal ring.

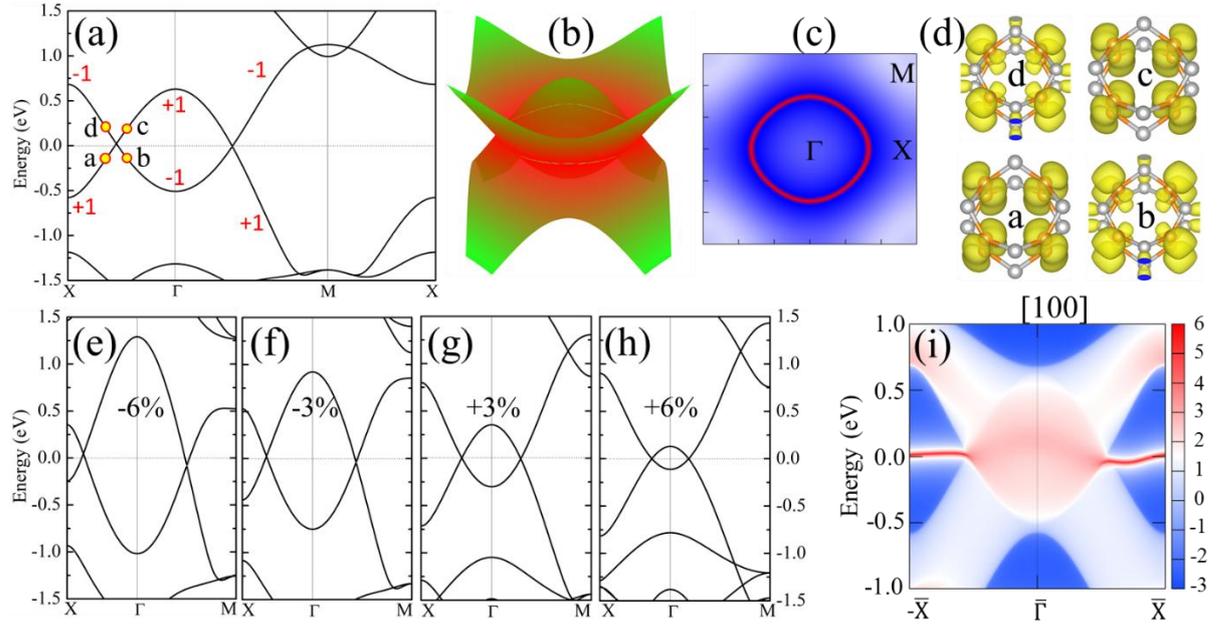

FIG. 6. (Color online) Electronic structure and topological properties of 2D-(4,0)-$C_{16}$. (a) Band structure of 2D-(4,0)-$C_{16}$ along the high-symmetry paths in the BZ, in which "-" and "+" represent the horizontal mirror eigenvalues. (b) 3D energy bands of 2D-(4,0)-$C_{16}$ near the Fermi level. (c) The distribution of the nodal ring (the red) surrounding point Γ in the BZ for 2D-(4,0)-$C_{16}$. (d) Charge density distributions at four points (a, b, c, d) in Fig. 6(a). (e-h) The band

structures of 2D-(4,0)-$C_{16}$ under -6%, -3%, +3%, and +6% in-plane biaxial strains, in which the negative and positive signs represent compression and tension strains, respectively. (i) The edge states of semi-infinite 2D-(4,0)-$C_{16}$ along the direction of [100].

On the basis of 2D-(4,0)-$C_{16}$ [see Fig. 3(e)], if it is further extended along the *c*-axis, a 3D carbon network structure can be obtained only by the (4,0)-CNT interconnections, namely 3D-(4,0)-$C_{16}$ [see Fig. 3(h)]. The band structure of 3D-(4,0)-$C_{16}$ is shown in Fig. 7(a), where we can observe some band crossing points along high-symmetry paths near the Fermi level. By calculating the irreducible representation, we find that the two bands involved in these crossing points on the Γ-Z-R and A-Z-X paths come from different irreducible representations, indicating that these crossed points are protected by crystalline symmetry. However, the two intersecting bands on the M-Z path belong to the same irreducible representation, so they hybridize with each other to open a small gap (21 meV), which is also confirmed by our calculations [see Fig. 7(b)]. In addition, these bands near the Fermi level are mainly the contribution of graphene-like $p_z$ orbitals from C2 atoms [see Fig. S5 in the SM and Fig. 7(d)], and the structure exactly matches our TB model when the remaining $sp^3$ atoms (C1) are removed. It is worth noting that 3D-(4,0)-$C_{16}$ has inversion (*P*) and time-reversal (*T*) symmetries, and these crossing points are not isolated for a *PT*-symmetric system without SOC. Indeed, a careful search reveals that these gapless band crossing points form a nodal cage around the Z point in the whole BZ [see Fig. 7(c)]. To confirm the existence of a "drumhead-like" surface state in the nodal cage, we investigated the surface spectrum of 3D-(4,0)-$C_{16}$ on the (001) surface [see Figs. 7(e-f)]. Taking the nodal ring on the $k_z = \pi$ or -$\pi$ plane as an example, we fix the energy at the intersection of the A-Z path to obtain the surface spectrum for the (001) surface, from which we can find that the "drumhead-like" surface state is nestled within the projected nodal ring [see Fig. 7(f)].

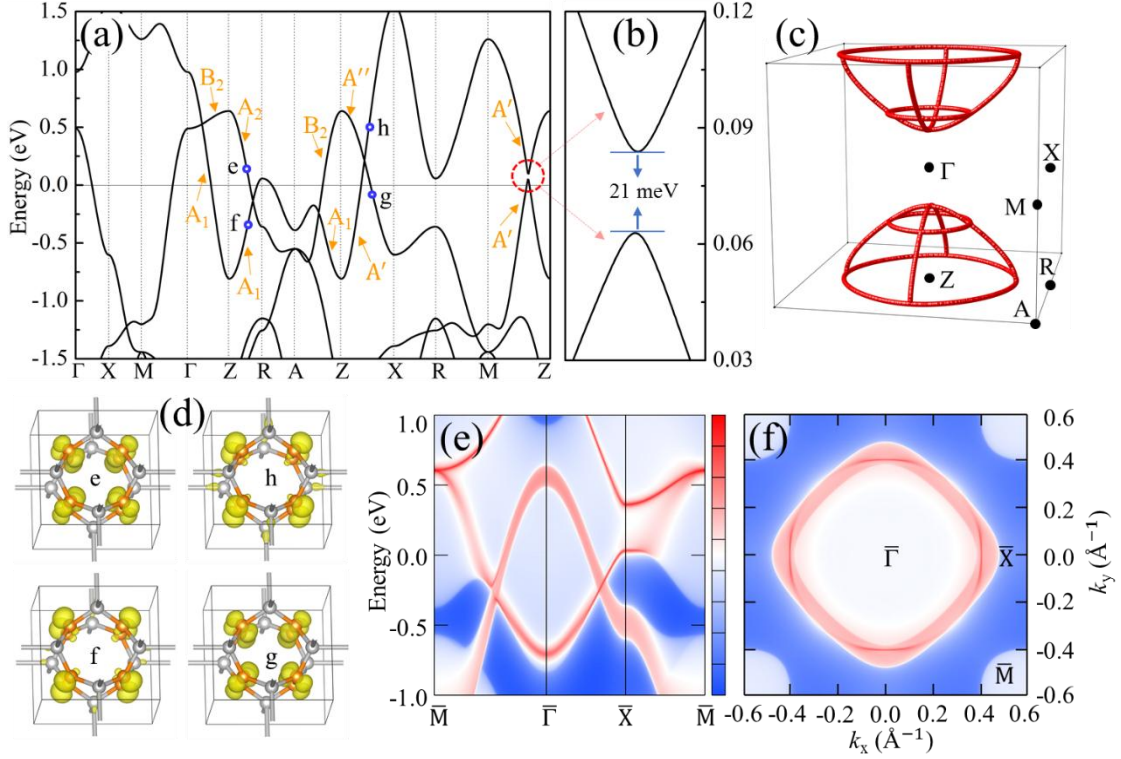

FIG. 7. (Color online) Electronic structure and topological properties of 3D-(4,0)-$C_{16}$. (a) Electronic band structure of 3D-(4,0)-$C_{16}$. The irreducible representations of two intersecting bands along high-symmetry paths near the Fermi level are also indicated. (b) Enlarged band structure near the seemingly intersections on the M-Z path in Fig. 7(a). (c) The nodal cage composed of all zero-gap nodal points formed by two intersecting bands near the Fermi level in the first BZ. (d) Charge density distributions at four points (e, f, g, h) in Fig. 7(a). (e) Surface energy bands of 3D-(4,0)-$C_{16}$ along the high-symmetry paths in the projected 2D BZ of the (001) surface. (f) Surface spectrum of 3D-(4,0)-C16 on the (001) surface at a fixed energy $E = E_f - 0.238$ eV. The energy is the one at the intersection of the A-Z path in Fig. 7(a).

Before concluding, we would like to discuss the possible advantages and potential applications of (4,0)-CNT derivatives. To the best of our knowledge, there is rarely a typical material system that can be used as an ideal platform to study the topological states and topological phase transitions as the system transforms from 1D nanoribbons to 2D sheet then to 3D bulk[57]. In this work, we demonstrate that the structurally stable (4,0) carbon nanotube derivatives are an ideal platform to explore the semiconductor/nodal-point semimetal states in 1D nanoribbons [1D-(4,0)-$C_{16}H_4$ and

1D-(4,0)-$C_{32}H_4$], nodal-ring semimetal state in 2D sheet [2D-(4,0)-$C_{16}$], and nodal-cage semimetal state in 3D bulk [3D-(4,0)-$C_{16}$]. Importantly, (4,0)-CNT derivatives are only composed of (4,0)-CNT as the building block, while (4,0)-CNT has been observed experimentally[58]. With the rapid development of experimental technology, it is believed that 3D-(4,0)-$C_{16}$ can be obtained experimentally by squeezing (4,0)-CNTs or by the bottom-up synthesis method[59]. In addition, 3D-(4,0)-$C_{16}$ has a large cavity and a transport channel along the nanotube direction, which may be conducive to improving the reversible storage space and to the capacity of hydrogen or lithium atoms.

## Conclusion

To summarize, we propose a tetragonal TB model composed of eight sites based on the upper and lower two-layer squares. With the change of system dimensions, the model demonstrates the complete topological phase transitions from a trivial insulator to various nontrivial topological semimetals, including nodal-point semimetal, nodal-ring semimetal, and nodal-cage semimetal. Following the guidance of this TB model, we propose a series of (4,0)-CNT derivatives [such as 1D-(4,0)-$C_{16}H_4$, 1D-(4,0)-$C_{32}H_4$, 2D-(4,0)-$C_{16}$, and 3D-(4,0)-$C_{16}$] that can serve as an ideal material platforms for realizing multiple topological states and topological phase transitions. Our work not only provides a new exceptional member of the 2D and 3D topological carbon material family, but also establishes a typical template for exploring topological phase transitions with dimension evolution.

## Author contributions

Y.G. conceived and designed the project, carried out the TB model and DFT calculations, and writing-original draft preparation. Y.D. and Y.Y.B. investigation, Methodology, and Software, data curation. W.W. and Q.W. analyzed the results, writing-reviewing, and editing. Y.G, K.L., Y.L., and Z.-Y.L. writing-reviewing, conceptualization, supervision, and project administration.

## Competing interests

## Acknowledgements

We wish to thank Chengyong Zhong for helpful discussions. This work was supported by the National Natural Science Foundation of China (Grants No. 12304202, No. 11934020, and No. 12174443), the National Key R&D Program of China (Grant No. 2019YFA0308603), the Innovation Capability Improvement Project of Hebei Province (Grant No. 22567605H), the Natural Science Foundation of Hebei Province (Grant No. A2023203007), and the Special Funding in the Project of Qilu Young Scholar Program of Shandong University. Y. G is also grateful for the support of HZWTECH for providing computational facilities.